\documentclass[]{emulateapj}
\usepackage{natbib}
\usepackage{xspace}
\usepackage{enumitem}
\usepackage{amsmath}
\usepackage{booktabs}

\newcommand{\ergs}{$ergs^{-1}$\xspace}

\newcommand{\chandra}{{\it Chandra}\xspace}
\newcommand{\hst}{{\it HST}\xspace}

\def\Msun{\hbox{$\thinspace M_{\odot}$}\xspace}

\def\LKsun{\hbox{$\thinspace L_{K\odot}$}\xspace}

\def\lk{\hbox{$\thinspace L_{K}$}\xspace}
\def\lx{\hbox{$\thinspace L_{x}$}\xspace}

\shorttitle{The XLF of LMXBs in the field, metal-rich and metal-poor GCs}
\shortauthors{Peacock et al.}

\begin{document}

\title{The X-ray luminosity function of low mass X-ray binaries in early-type galaxies, their metal-rich, and metal-poor globular clusters}

\author{Mark B. Peacock and Stephen E. Zepf}
\affil{Department of Physics and Astronomy, Michigan State University, East Lansing, MI 48824, USA}
\email{MBP: mpeacock@msu.edu}

\begin{abstract}
\label{sec:abstract}

We present the X-ray luminosity function (XLF) of low mass X-ray binaries (LMXBs) in the globular clusters (GCs) and fields of seven early-types galaxies. These galaxies are selected to have both deep \chandra observations, which allow their LMXB populations to be observed to X-ray luminosities of $10^{37}-10^{38}$\ergs, and \hst optical mosaics which enable the X-ray sources to be separated into field LMXBs, GC LMXBs, and contaminating background and foreground sources. We find that at all luminosities the number of field LMXBs per stellar mass is similar in these galaxies. This suggests that the field LMXB populations in these galaxies are not effected by the GC specific frequency, and that properties such as binary fraction and the stellar initial mass function are either similar across the sample, or change in a way that does not effect the number of LMXBs. We compare the XLF of the field LMXBs to that of the GC LMXBs and find that they are significantly different with a p-value of $3\times10^{-6}$ (equivalent to 4.7$\sigma$ for a normal distribution). The difference is such that the XLF of the GC LMXBs is flatter than that of the field LMXBs, with the GCs hosting relatively more bright sources and fewer faint sources. A comparison of the XLF of the metal-rich and metal-poor GCs hints that the metal-poor GCs may have more bright LMXBs, but the difference is not statistically significant.

\end{abstract}


\section{Introduction}
\label{sec:intro}

Low mass X-ray binaries (LMXBs) are systems in which a black hole or neutron star accretes gas from a low mass companion star. These systems dominate the X-ray point source population in the old stellar populations of early-type galaxies. In the \chandra era, it is possible to resolve the LMXB populations of local galaxies. However, exploiting the full potential of such observations requires similarly high resolution \hst observations to identify optical counterparts. This enables X-ray sources to be separated into field and GC LMXBs and contamination from background AGN to be removed, by detecting their host galaxies. 

Globular clusters (GCs) are highly efficient at forming LMXBs, as seen in the Milky Way \citep[e.g.][]{Clark75, Liu01, Pooley03} and local early-type galaxies, where 20-70$\%$ of the LMXBs observed reside in GCs \citep[e.g.][]{Angelini01,Kundu02,Jordan04}. This is thought to be the result of dynamical formation of LMXBs in the dense cluster cores, a theory supported by an observed correlation between the presence of LMXBs and stellar interaction rate \citep{Verbunt87, Jordan07, Peacock09}. However, in the field of a galaxy, stellar densities are typically too low for LMXBs to form dynamically. Instead they must either form dynamically in GCs and then be ejected into the galaxy \citep{Grindlay84}, or from the evolution and tightening of a primordial binary system \citep[e.g.][and references therein]{Ivanova13}. 

Observations of GC LMXBs in both the Milky Way \citep{Grindlay93} and M~31 \citep[][]{Bellazzini95, Peacock10b} show that they favor higher metallicity environments. This trend is strongly confirmed in \chandra observations of elliptical galaxies where the fraction of GCs hosting LMXBs is around three times higher in the red (metal-rich) clusters than the blue (metal-poor) clusters \citep[e.g.][]{Kundu07, Sivakoff07, Kim13}. 

Different formation mechanisms of LMXBs in the field, metal-rich and metal-poor GCs may produce differences in their donor stars, primaries (i.e. neutron stars or black holes) and orbital parameters. These differences will effect the luminosity of the LMXBs and hence careful measurement of the X-ray luminosity function (XLF) gives clues to the nature of the LMXB population and can be used to test for differences between environments. 

Previous studies have considered the XLF of LMXBs with varying sample sizes, detection limits, and contamination. \citet{Humphrey08} combined shallow \chandra observations with WFPC2 optical observations to study bright LMXBs in the central regions of a sample of early-type galaxies. They found the GC and field XLFs to be well represented by a double powerlaw, with a break at $2.2\times10^{38}ergs^{-1}$. In deeper observations of Cen~A \citep[][]{Voss09} and NGC~3379, NGC~4278, NGC~4697 \citep[][]{Kim09} the XLFs of the GC LMXBs were found to be significantly flatter at low luminosities. \citet{Zhang11} used a heterogeneous sample of local group and nearby galaxies to propose that this difference may extend throughout the XLF. Recently, \citet{Kim13} tested for differences in the LMXB populations of metal-rich and metal-poor GCs, finding no significant differences between the populations. 

\begin{table*}
 {\centering
  \caption{Galaxy data \label{tab:galaxy_data}}
  \begin{tabular}{@{}r@{\hskip 4mm}crrclccrrc@{\hskip 4mm}ccc@{}}  
  \hline 
  \hline
  \noalign{\vskip 1mm}    
  & \multicolumn{10}{c}{Galaxy data} & \multicolumn{3}{c}{\chandra data$^{{\rm viii}}$}\\ \cmidrule(r{15pt}){2-11}\cmidrule(){12-14}
 NGC     & distance$^{\rm i}$ & $S_{N}^{\rm ii}$ & $T^{\rm ii}$ & B-V$^{\rm iii}$  & ${\sigma_{\rm 1kpc}}^{\rm iv}$ & $r_{inner}^{\rm v}$ & $r_{ext}^{\rm vi}$ & $e^{\rm vi}$ & $L_{K}^{\rm vi}$(cover)  & $M/\lk^{\rm vii}$ &  exp. time &     90$\%$ limit             & ref.  \\ 
      &  (Mpc)  &   &  &   &  (kms$^{-1}$)   & (arcsec) & (arcsec) & &  (10$^{10}$\LKsun)  &  &   (ksec)  & (10$^{38}$ergs$^{-1}$)  &       \\ [1ex]
   \hline
   1399  &  20.0 & 3.6 & 15.8  & 0.95 & 279.9 & 10   & 202.2 & 0.00  &  23.1 (18.0)  & 0.85 & 101  & 1.20  & 1   \\
   3379  &  10.6 & 0.5 &   5.0   & 0.94 & 196.8  & 10   & 191.7 & 0.15  &   5.5  (4.2)  & 0.83 &  324  & 0.05  & 2   \\
   4278  & 16.1 &  3.6 &  30.1  & 0.90 & 228.0 & 10   & 155.0 & 0.07  &   6.9   (4.5)  & 0.78 &  458  & 0.10  & 3   \\
   4472  & 16.7  & 1.1 &   6.1  & 0.95 & 288.4 & 15   & 313.4 & 0.19  &  37.7 (20.3) & 0.85 & 380  & 0.90  & 4   \\
   4594  &  9.0  &  2.1 &  11.4  & 0.84 & 251.2 & 22.5*& 297.1 & 0.46 & 17.2   (7.2) & 0.72 & 174  &  0.20 & 5   \\
   4649  & 16.5  & 3.8 &  12.0  & 0.95 & 307.6 & 15   & 241.3 & 0.19  & 27.3 (16.7) & 0.85 &  300  &  0.60 & 6   \\
   4697  & 11.7  & 1.5 &  11.1  & 0.89 & 180.3 & 10   & 240.2 & 0.37  &   7.4   (6.0) & 0.77 & 132  &  0.14 & 7   \\
   \hline
   \end{tabular}\\
 }
\footnotesize
\label{tab:galaxy_data}
\vspace{1mm}
$^{{\rm i}}$distances from surface brightness fluctuation measurements \citep{Blakeslee01, Jensen03, Blakeslee09}; 
$^{{\rm ii}}$GC specific frequency \citep[$S_{N}$;][]{Kundu01, Rhode01, Kim09, Paolillo11, Mineo14} and GC mass specific frequency ($T$; See Section \ref{sec:optical}); 
$^{{\rm iii}}$dereddened B-V color of galaxy \citep{deVaucouleurs91}; 
$^{{\rm iv}}$galaxy's velocity dispersion \citep{Saglia00, Jardel11, Cappellari12};   
$^{{\rm v}}$The radius defining the central region that is excluded from our analysis, *For NGC~4594 we remove an elliptical inner region with semi-minor axis = 22.5\arcsec and semi-major axis = 168\arcsec; 
$^{{\rm vi}}$Galaxy data from the two micron all sky survey (2MASS) large galaxy atlas (LGA) \citep{Jarrett03}, `total' extrapolated galaxy semi-major axis (${\rm r_{ext}}$), K-band ellipticity ($e=1-b/a$), K-band luminosity of galaxy within this ellipse (\lk) and, in brackets, the K-band light covered by this study; 
$^{{\rm vii}}$The K-band stellar mass to light ratio ($M/\lk$) calculated from \lk and B-V using the correlation from \citet{Bell01}; 
$^{{\rm viii}}$total \chandra exposure times, 90$\%$ completeness limit, and references for the X-ray catalog: (1) \citet{Paolillo11}; (2) \citet{Brassington08}; (3) \citet{Brassington09}; (4) \citet{Joseph13}; (5) \citet{Li10}; (6) \citet{Luo13}; (7) \citet{Sivakoff08} \\ 
\end{table*}

In \citet{Peacock14}, we recently presented an analysis of the LMXB populations of seven early-type galaxies within 20~Mpc. This dataset focusses on producing clean catalogs of field and GC LMXBs across these galaxies. To do this, we require that the galaxies have deep \chandra data and exclude the crowded central regions from our analysis. All of the galaxies have \hst optical mosaics that allow us to extend the study of their LMXB populations out to their D25 ellipse. Through careful alignment of these images with these \chandra data we can produce clean samples of field and GC LMXBs. We also reliably detect the optical counterparts to background AGN. This allows us to directly remove these sources, rather than rely on statistical corrections which introduce additional uncertainties. 

In this paper, we present an analysis of the XLF based on this sample of old early-type galaxies: Section \ref{sec:data} reviews the data and counterpart matching/classification; Section \ref{sec:xlf} presents the observed XLFs of these galaxy's field LMXBs, GC LMXBs, and red/blue GC LMXBs; Section \ref{sec:xlf_field} discusses the similarity of the field LMXBs in these galaxies; while in Sections \ref{sec:xlf_field_gcs} and \ref{sec:xlf_gcr_gcb} consider differences between the XLF of field and GC LMXBs and between red and blue GC LMXBs, respectively.

\section{Data}
\label{sec:data}

Here, we review the dataset analyzed in this paper. We note that this was previously presented in \citet{Peacock14}, to which we refer the reader for full details. 

Our sample of galaxies and their X-ray data are summarized in Table \ref{tab:galaxy_data}. The galaxies are all early-type galaxies within 20~Mpc and have stellar masses of $(5-34)\times10^{10}\Msun$ \citep[as inferred from their K-band luminosity and assuming $M/\lk=0.9$;][]{Jarrett03, Fall13}. The galaxies are thought to have similar metallicities and (old) ages. We are not aware of a homogeneous study of the stellar populations of all of the galaxies in our sample. However, \citet{Sanchez-Blazquez06b} calculated the ages ($t$) and metallicities ($[Z/H]$) of 5/7 of our galaxies to be in the ranges $9.9<t<10.1$~Gyr and $0.08<[Z/H]<0.14$, respectively. We can also estimate the ages and metallicities in a statistical sense, by assuming they follow the well established correlations observed for early-type galaxies, that $log(t)=-0.11+0.47log(\sigma)$ and $[Z/H]=-1.34+0.65 log(\sigma)$ \citep{Thomas10}. For the range of $\sigma$ covered by these galaxies, this implies ages of 8.9--11.5~Gyr and metallicities of 0.12--0.27, broadly consistent with those calculated from the stellar populations models and again implying similar populations. 


\begin{figure*}
 \centering
 \includegraphics[width=170mm,angle=0]{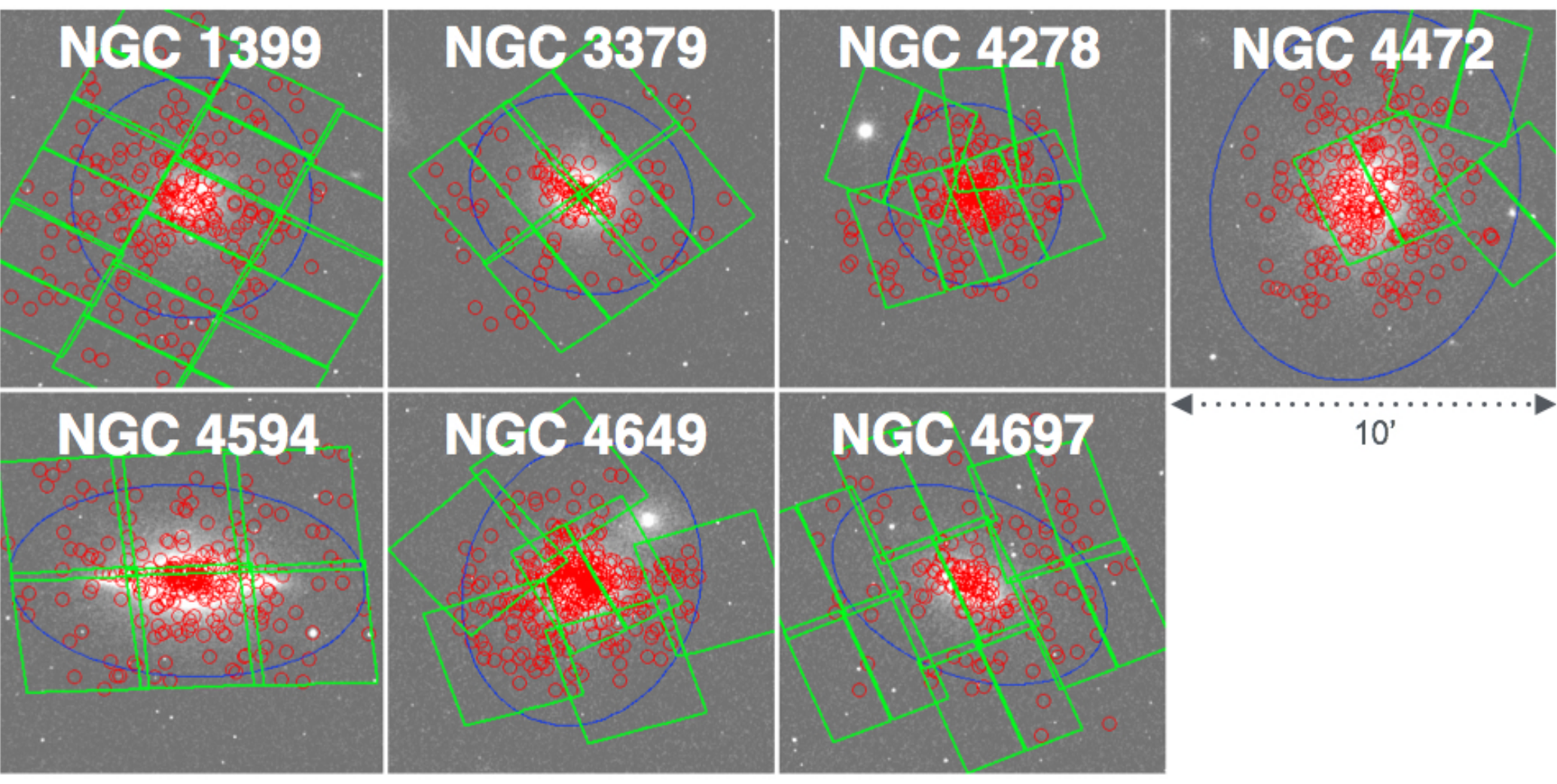} 
 \caption{K-band images of the seven galaxies in our sample from the two micron all sky survey (2MASS). The blue lines show the $r_{k,ext}$ ellipse for each galaxy from the 2MASS/LGA \citep{Jarrett03}, this is taken as the outer boundary for X-ray sources to be considered part of the field of the galaxy. The red points show the locations of X-ray point sources. The green lines show the locations of the \hst/ACS optical images. \\}
 \label{fig:fields} 
\end{figure*}

\subsection{X-ray data} 
\label{sec:data_xray}

As shown in Table \ref{tab:galaxy_data}, these galaxies have deep \chandra observations of $100-450$~ksec, resulting in 90$\%$ detection limits of $(0.05-1.2)\times10^{38} ergs^{-1}$. Based on these data, X-ray source catalogs have previously been published in studies of the individual galaxies. We utilize these previous datasets, from the works cited in Table \ref{tab:galaxy_data}. A more homogeneous catalog is available from \citet{Liu11}, but this is found to be less complete than the individual studies, so is not utilized. However, it is used to: convert the quoted flux of sources in NGC~1399's catalog from $photon/s$ to $erg/s$; scale the \lx values for sources in NGC~4472 to be consistent with \citet{Liu11} and previous studies; and to ensure there are no systematic offsets in the \lx of sources in the other galaxies in our sample. We also confirm that the detection limits quoted by these different studies are consistent with those predicted by the web-based simulator {\sc pimms} (v4.6a)\footnote{http://cxc.harvard.edu/toolkit/pimms.jsp}.

We wish to study the cleanest population of LMXBs possible. We  therefore restrict our analysis to only X-ray sources with $\lx$ greater than the 90$\%$ detection limit and to those sources that have galactocentric radii ($r$) in the range  $r_{inner}<r<r_{ext}$. The values of these radii are quoted in Table \ref{tab:galaxy_data}. $r_{ext}$ is the ellipse defined by the 2MASS LGA \citep[and is similar to the optical D25 radius;][]{Jarrett03}. $r_{inner}$ defines the central circular region that is excluded from our analysis. Inside of this radius crowding and hot gas emission result in larger errors on \lx, higher detection limits and difficulty with optical counterpart association.  

Figure \ref{fig:fields} shows 2MASS K-band images of the seven galaxies in our sample. The blue ellipses indicate the $r_{ext}$ region, at which we truncate our analysis. All of the X-ray sources detected in these galaxies are shown as red points. The green regions show the locations of the \hst/ACS images used to detect optical counterparts (discussed below). 

\subsection{Optical data/ counterparts} 
\label{sec:optical}

\begin{figure*}
 \centering
 \includegraphics[width=176mm,angle=0]{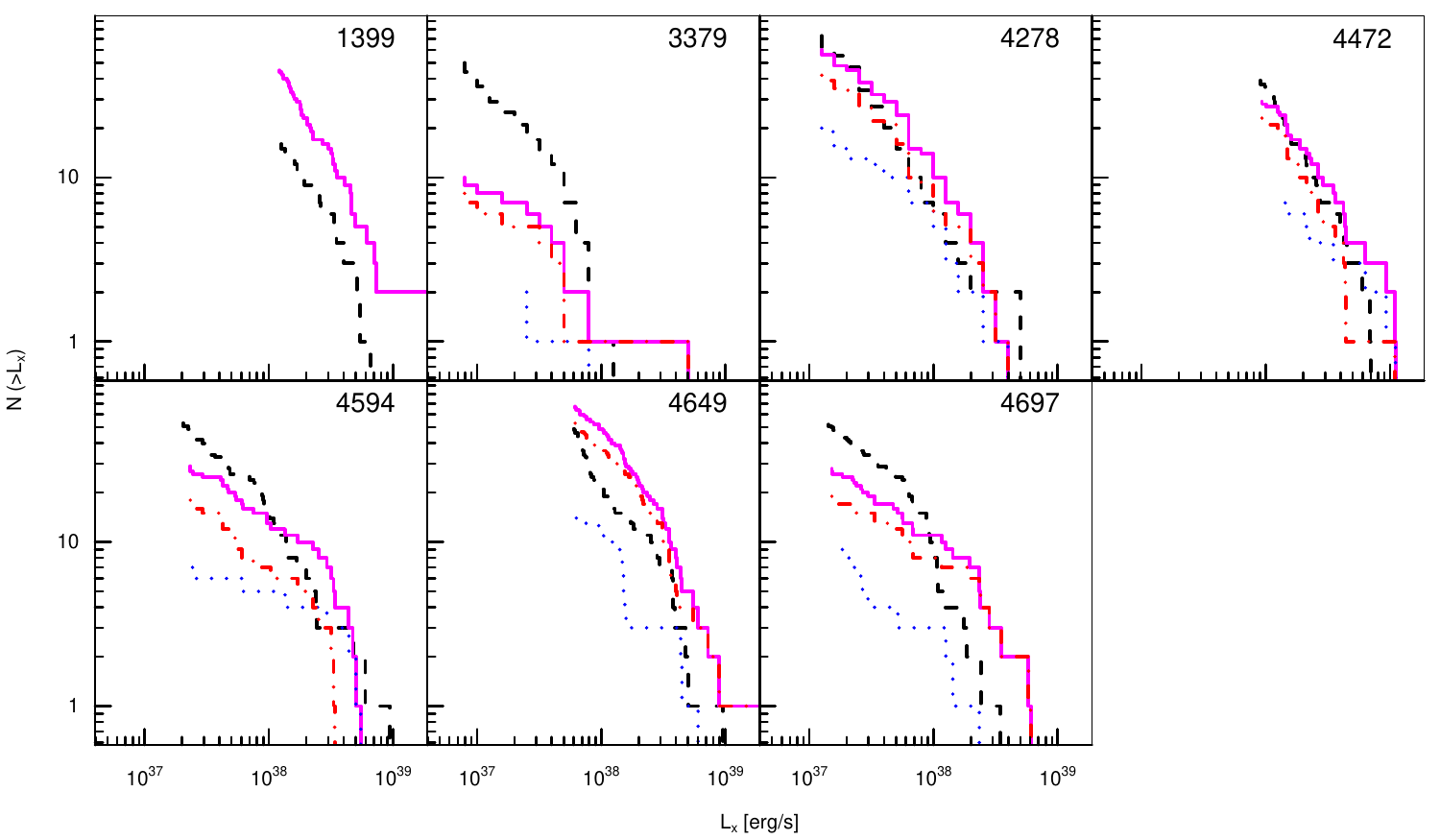} 
 \caption{The XLF of the LMXBs observed in each galaxy. The black-dashed line shows the field LMXB population. The magenta-solid line shows the GC LMXB population. The red-dot-dashed and blue-dotted lines split the GC LMXBs into metal-rich and metal-poor populations, respectively. \\}
 \label{fig:xlf_multi} 
\end{figure*}

Our sample is selected to have \hst/ACS mosaics which cover a large fraction of the galaxies out to $r_{ext}$ and are deep enough to detect their GCs and background AGN. To ensure that counterparts are reliably detected, we only study galaxy regions covered by these images. For NGC~1399, the mosaic ACS image is only available through one filter. This is sufficient to identify GCs and background galaxies, but does not enable us to study the colors of the GCs and so is not utilized in our latter analysis of red/blue GCs. The mosaic of NGC~4594 is through the F435W and F625W filters (equivalent to B and $r$). For all of the other galaxies the mosaic images are through the F475W and F850LP filters (equivalent to $g$ and $z$). 

We use the pipeline produced images available from the HLA\footnote{Hubble Legacy Archive: http://hla.stsci.edu/} and subtract the smooth galaxy light from these images using a ring median filter (with an inner radius of 30 pixels). Each image is aligned to the X-ray source catalog by using the {\sc iraf} task {\sc tfinder} to update the images world coordinate system (WCS). We then use {\sc sextractor} to perform photometry through a 0.25$\arcsec$ aperture. Aperture corrections are applied from $0.25-0.5\arcsec$ based on an empirical estimate from {\sc sextractor} photometry of bright GC like sources. A further correction is applied from 0.5$\arcsec$ to infinity based on the point source corrections published by \citet{Bohlin11}. 

From this final optical catalog, sources are classified as GC candidates if they: are extended, with {\sc sextractor} stellarity flag $<0.9$; are not too extended to be a cluster, with $\Delta mag(0.25-0.5\arcsec)<0.4$; have absolute magnitudes in the range $-12.0<z<-6.5$ ($-11.5< r<-6.5$ for NGC 4594); and have colors in the range $0.6<g - z<1.7$ ($0.7<B-r<1.5$ for NGC 4594). Our GC candidates show the clear bimodality observed previously in these (and other) galaxies. We split the clusters into red and blue GCs using the KMM method \citep{Ashman94}, implemented using the {\sc GMM} code of \citet{Muratov10}. This fits two equal width Gaussians to the GCs and assigns them to a red and a blue population. We run this fitting independently for each galaxy's GCs, allowing different widths and color cuts between the two populations. 

We also use these data to calculate the local GC mass specific frequency, $T=N_{\rm GCs}/(M_{G}/10^{9}M_{\odot})$. Here, $N_{\rm GCs}$ is the total number of GCs and is calculated by fitting the derived GC luminosity functions to a normal distribution with fixed parameters $\sigma=0.25$ and $M_{TO,z}=-8.45$ \citep[consistent with previous studies, e.g.][]{Villegas10}. $M_{G}$ is the total stellar mass and is calculated from \lk and $M/\lk$, as listed in Table \ref{tab:galaxy_data}. We do not extrapolate these values to find a total $T$ for each galaxy, rather we quote this local value which is appropriate for comparison to our X-ray analysis. 

The X-ray source catalog is matched to this optical catalog based on the aligned WCS locations using {\sc stilts} \citep{Taylor06}. To produce clean samples of field and GC LMXBs we take conservative matching radii. Field LMXBs are taken to be X-ray sources with no optical counterparts within 0.8\arcsec. GC LMXBs have to be within 0.4\arcsec of a GC candidate. Examination of the offset between the matched sources, and false matches produced from shifted source locations, suggests that both samples should have little contamination present. The number of X-ray sources with non-cluster optical counterparts is consistent with that expected from background AGN \citep{Kim07,Peacock14}. 

\newpage
\section{Observed XLFs} 
\label{sec:xlf}

Figure \ref{fig:xlf_multi}, shows the XLFs of each galaxy's LMXBs in the field (black-dashed lines), all of its GCs (purple-solid lines), its metal-rich GCs (red-dot-dashed lines) and its metal-poor GCs (blue-dotted lines). This figure suggests there may be differences in the XLFs among the different populations in some of these galaxies. To test for the significance of possible differences between the XLFs, we run both Wilcoxon rank sum and Anderson-Darling tests over the data (implemented using {\sc R}'s {\sc ksamples} package). We quote only the p-values from the Anderson-Darling tests, but note that values derived from both statistics are consistent. 

\begin{figure}
 \centering
 \vspace{5mm}
 \includegraphics[width=86mm,angle=0]{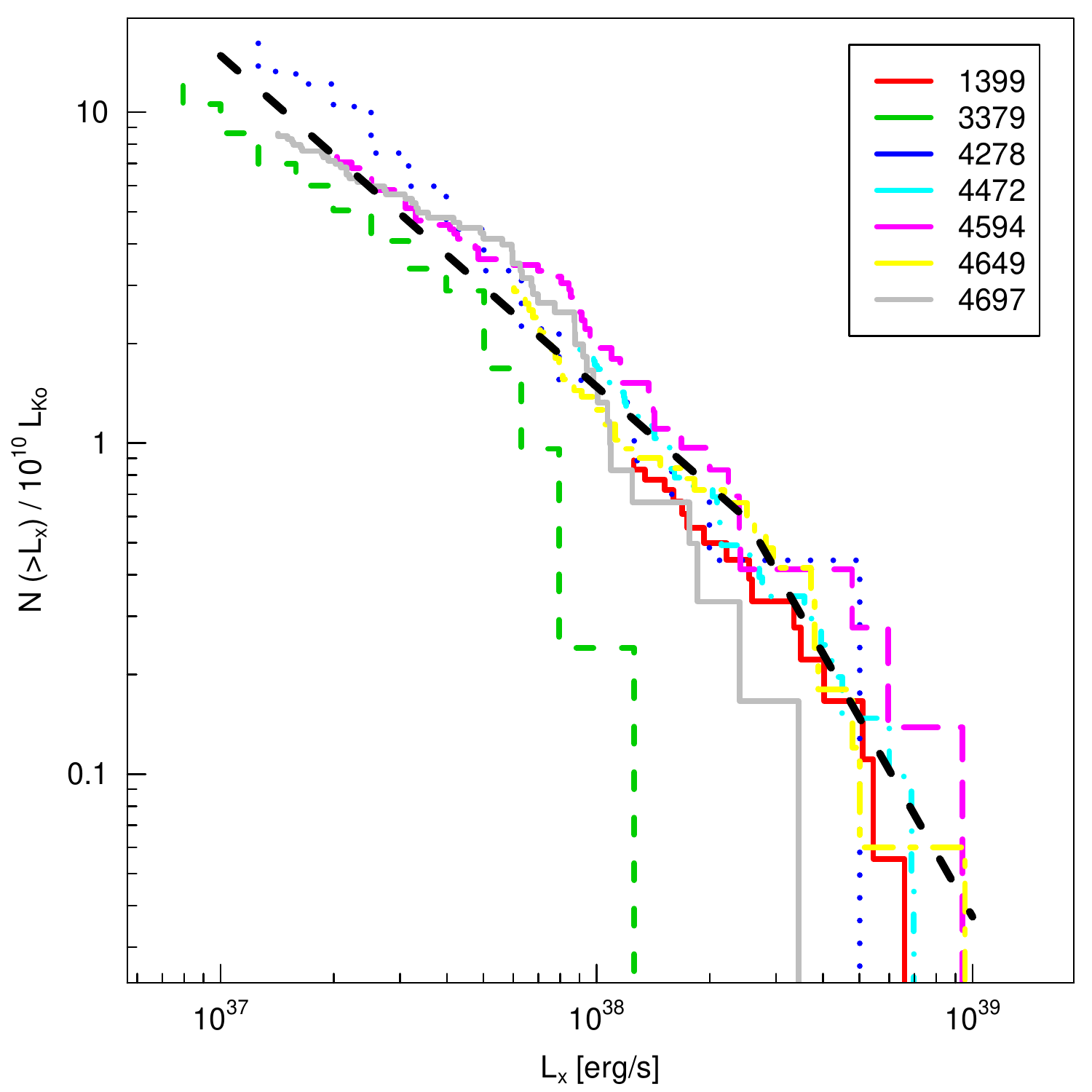} 
 \caption{The XLFs of LMXBs in the fields of the seven galaxies studied. These XLFs are scaled by the stellar K-band light covered. The black dashed line shows a broken power-law representation of these data. The small offsets between these XLFs is indicative that these galaxies form a similar number of LMXBs per stellar mass.  \\}
 \label{fig:xlf_sc_lum} 
\end{figure}

When analyzing each galaxy on its own, only NGC~4649 is found to have a highly significant difference between its field and GC LMXBs, with a p-value of $4.4\times10^{-5}$ (equivalent to $4.0\sigma$ for a normal distribution). The other galaxies show a similar difference, with the GCs hosting relatively more of the brightest LMXBs and less of the faintest. However, for these galaxies, this difference is not significant. 

No significant differences are observed between the metal-rich and metal-poor GC LMXBs in any of the galaxies. The metal-poor GC LMXBs in NGC~4472 appear to be brighter than their metal-rich counterparts but, with a p-value of only 0.054. 

In the following sections, we consider the similarity of the XLF of field LMXBs in these early-type galaxies (Section \ref{sec:xlf_field}) and utilize the improved statistics from the combined dataset to investigate differences in the XLF of field vs. GC LMXBs (Section \ref{sec:xlf_field_gcs}) and metal-rich vs. metal-poor GC LMXBs (Section \ref{sec:xlf_gcr_gcb}).

\section{The field LMXB populations}
\label{sec:xlf_field}

\begin{figure}
 \centering
 \includegraphics[width=90mm,angle=0]{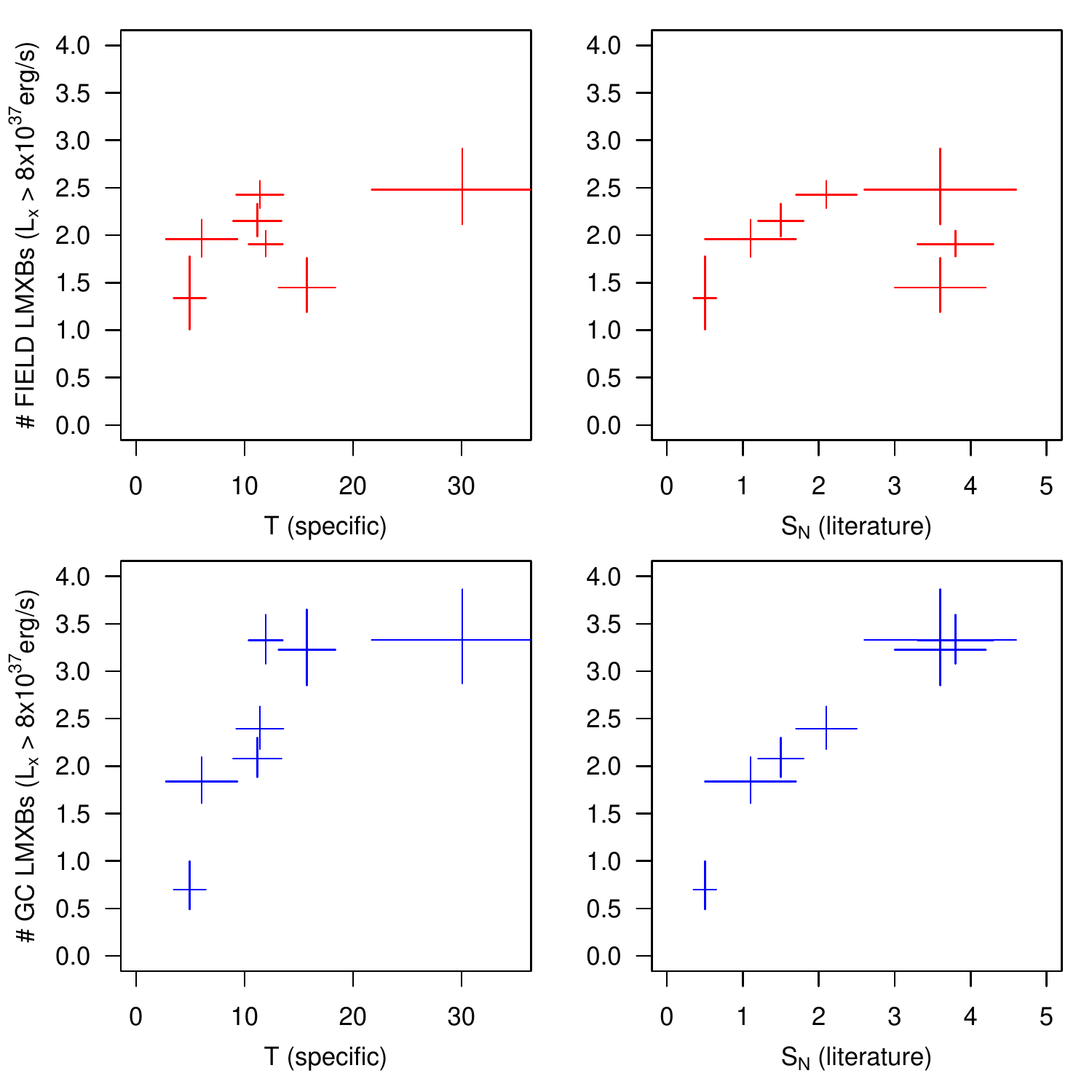} 
 \caption{The number of LMXBs/\lk in the field (top, red) and GCs (bottom, blue) as a function of the local GC mass specific frequency (T, left) and specific frequency estimated from previous studies ($S_{N}$, right). The GC specific frequency appears to have little influence of the field LMXB population. The correlation between $S_{N}$ and GC LMXBs is simply due to the larger number of GCs present per \lk for higher $S_{N}$.}
 \label{fig:N_Sn} 
\end{figure}

Figure \ref{fig:xlf_sc_lum} shows the XLF for LMXBs in the fields of the seven early-type galaxies in our sample, scaled by the total stellar light covered, N(LMXBs)/$10^{10}$\LKsun. We measure stellar light covered directly from 2MASS LGA images after carefully masking them to only include the galaxy regions covered by this study. It can be seen that for six of the seven galaxies observed, the XLFs are similar. This implies that the formation and evolution of LMXBs in these old stellar populations is similar. One galaxy, NGC~3379 has a lower number of LMXBs at a given \lx, although we note that this is exacerbated in Figure \ref{fig:xlf_sc_lum} by small number statistics at high \lx. This result was previously observed and discussed by \citet{Kim09} and \citet{Peacock14}. 

The dashed-black line in Figure \ref{fig:xlf_sc_lum} is a broken powerlaw representation of the XLF. It has the form:

\begin{equation}  \label{equ:bpl}
 N(>\lx) \propto \begin{cases}
    (\lx/2.5)^{-2.0}, & \text{if $\lx>2.5$}\\
    (\lx/2.5)^{-1.0}, & \text{otherwise}
 \end{cases}
\end{equation}
\\
where \lx is in units of $10^{38}$\ergs. The break luminosity and exponents are consistent with previously proposed values \citep[e.g.][]{Kim04,Humphrey08,Kim09,Kim10,Zhang12}. It can be seen that, over this range of \lx, this function provides a reasonable representation of the observed field XLF. We use this fit to each galaxy's XLF to estimate the number of field LMXBs with $\lx>8\times10^{37}$\ergs, per stellar light (\#LMXBs/\lk). We calculate the error on this value as the combination of the error on the normalization of the XLF and the variance observed when the XLF functions exponents are varied by $\pm0.2$ and the break luminosity by $\pm2$. Below, we consider whether the small variation in this number are correlated with any galaxy parameters.

\begin{figure}
 \centering
 \includegraphics[width=90mm,angle=0]{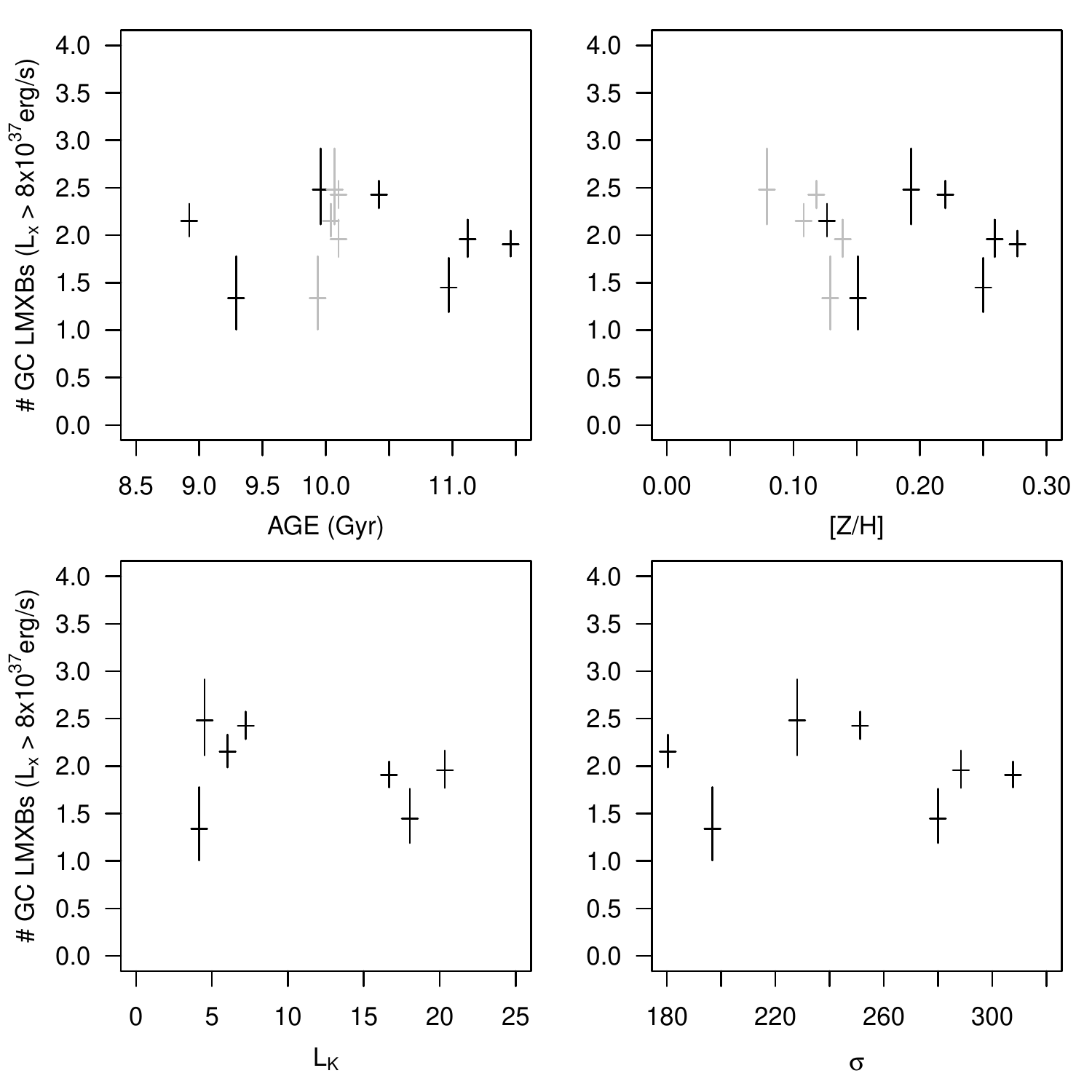} 
 \caption{The number of field LMXBs/\lk as a function of galaxy age (top left), metallicity ([Z/H], top right), \lk (bottom left), and velocity dispersion ($\sigma$, bottom right). Over the ranges covered by these galaxies, none of these parameters appear to influence the LMXB population. \\}
 \label{fig:N_gal_param} 
\end{figure}

\subsection{The specific frequency of GCs} 
\label{sec:Sn}

In the top panels of Figure \ref{fig:N_Sn}, we show the \#LMXBs/\lk as a function the local GC mass specific frequency, $T$ (see Table \ref{tab:galaxy_data} and Section \ref{sec:optical}). For comparison with previous studies, we also show the GC specific frequency, $S_{N},$ taken from the sources cited in Table \ref{tab:galaxy_data}. The GC specific frequency varies with galactocentric radii, so the local values should be more appropriate for comparison to our X-ray analysis. 

Previously, \citet{Kim09} proposed a potential correlation between  \#LMXBs/\lk and $S_{N}$, based on their study of NGC~3379, NGC~4278 and NGC~4697. This correlation remained, but was weaker with the addition of NGC~1399 and NGC~4649 \citep{Paolillo11, Mineo14}. Our analysis confirms these previous trends but the larger sample suggests that the correlation is even weaker, or not present. The observed lack of a correlation between the {\it field} LMXBs and $S_{N}$ suggests that an insignificant fraction of these LMXBs form in GCs. Sharper tests of this correlation would be enabled by new deep \chandra observations of galaxies with $S_{N}$ similar to the extreme values of NGC~3379 and NGC~4278. 

We note that other studies have shown significant correlations between  \#LMXBs/$\lk$ (or total LMXB \lx) and $S_{N}$ {\it without} distinguishing between field and GC LMXBs \citep[e.g.][]{Kim04, Boroson11, Zhang12}. However, such a trend will be driven by the higher number GCs present (and hence larger number of GC LMXBs) at higher $S_{N}$. We demonstrate this in the bottom panels of Figure \ref{fig:N_Sn}, where we show that there is a strong correlation between the number of GC LMXBs/$\lk$ and $S_{N}$. 

\subsection{Age} 

The top left panel of Figure \ref{fig:N_gal_param} shows that the age of the stellar populations appears to have little influence on the LMXB populations for these old galaxies. The models of \citet{Fragos08} suggest that there may be a large reduction in the number of LMXBs with increasing age and previous work has observed an increase in the number of LMXBs/\lk in galaxies of a few Gyr to galaxies of $8-10$~Gyr \citep{Kim10,Lehmer14}. However, our sample of galaxies have a relatively small spread in ages (estimated as $9-11.5$~Gyr, see Section \ref{sec:data}). For these old early-type galaxies, the similar LMXB populations observed are consistent with both this previous work and the models of \citet{Fragos08}. 

\begin{figure}
 \centering
 \includegraphics[width=90mm,angle=0]{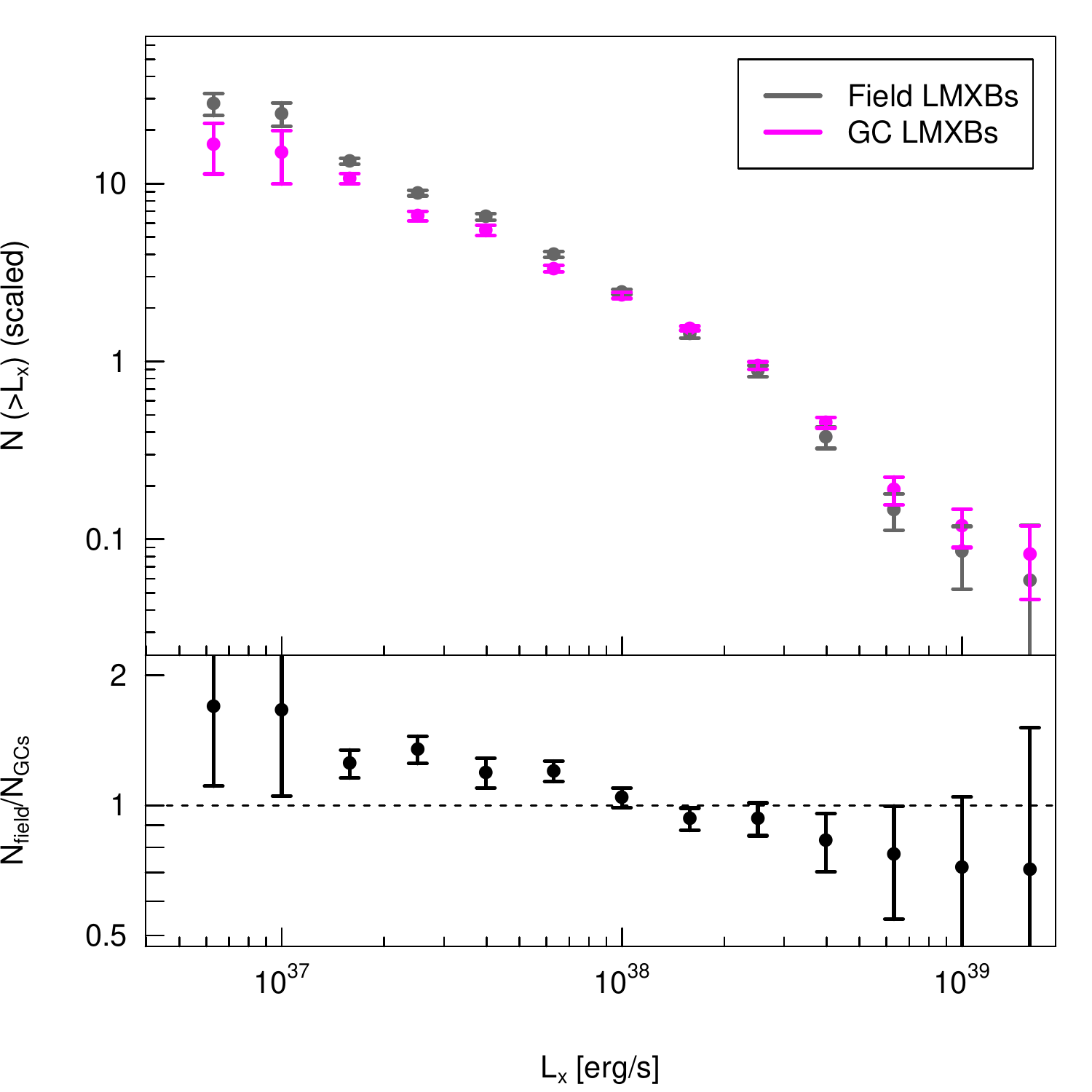} 
 \caption{(top panel): The average XLF of field (grey) and GC (magenta) LMXBs for all galaxies studied. (bottom panel): The ratio of field to GC LMXBs in each luminosity bin. The XLF of GC LMXBs is significantly different to that of the field, with relatively more bright sources and fewer faint sources. \\}
 \label{fig:xlf_field_gcs} 
\end{figure}

\begin{figure}
 \centering
 \includegraphics[width=90mm,angle=0]{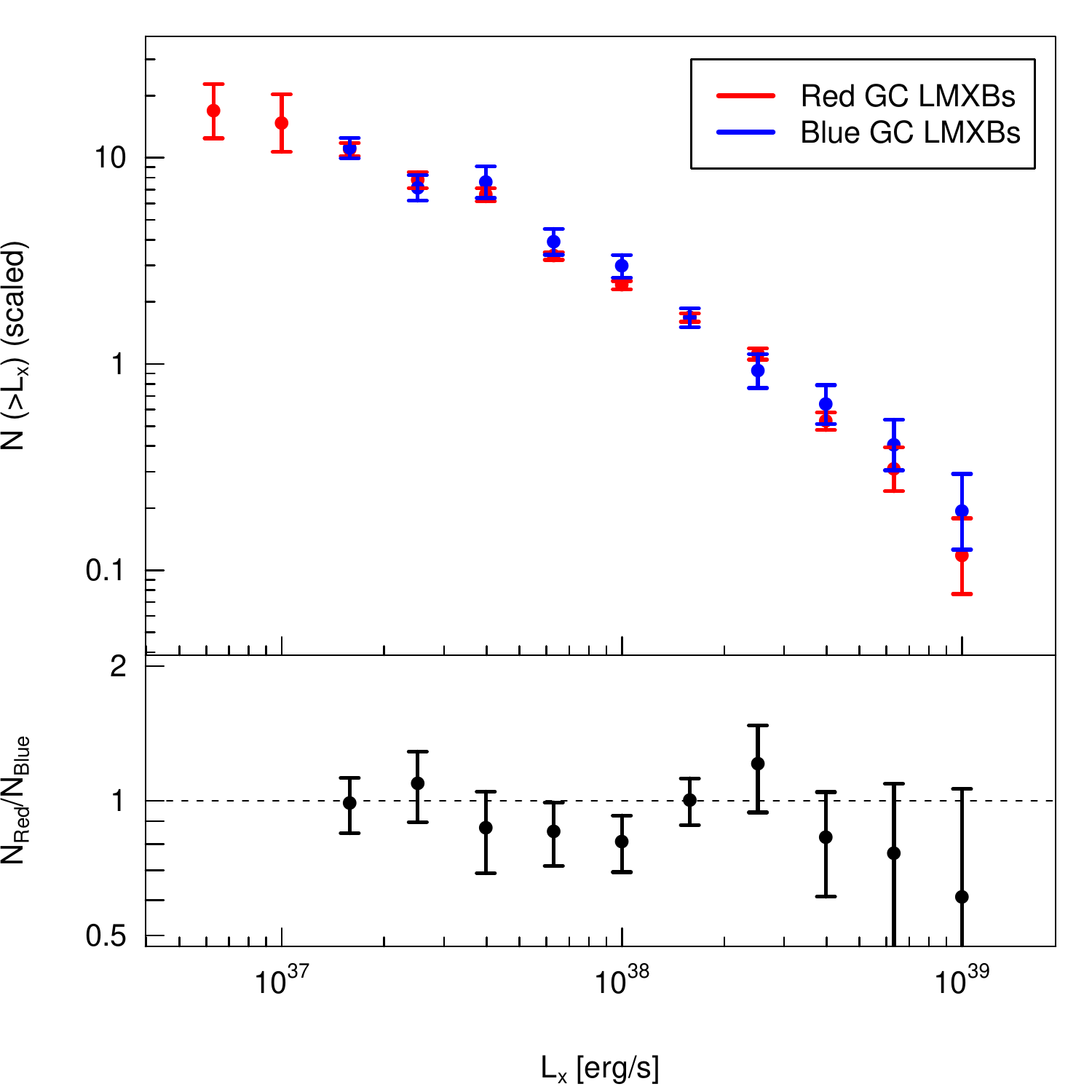} 
 \caption{(top panel): The XLF of LMXBs in red, metal-rich clusters (red points) and blue, metal-poor clusters (blue points). (bottom panel): the ratio of red to blue GC LMXBs in each luminosity bin. No significant difference is observed between the populations. \\}
 \label{fig:xlf_gcb_gcr} 
\end{figure}

\subsection{Metallicity} 

The top right panel of Figure \ref{fig:N_gal_param} shows that the small difference in the metallicity of these galaxies ($Z$) has little effect on the LMXB population. From observations of GC LMXBs, it is well established that the number of LMXBs present increases with metallicity (Z), such that $Z^{0.32}$ \citet{Sivakoff07}. However, GCs have a much broader spread in metallicity than these early-type galaxies (which we estimate to be in the range $0.12<Z<0.27$, see Section \ref{sec:data}). If this relationship applies to field LMXBs, it would imply a factor 1.3 increase from the most metal-poor to metal-rich galaxy. While no trend is observed in our data, this small correlation is consistent with the observations. 

\subsection{Galaxy mass and binary fraction} 

The bottom panels of Figure \ref{fig:N_gal_param}, show that there is also no correlation between \#LMXBs/$\lk$ and the K-band light and velocity dispersion of these galaxies. Such a trend may have been expected if the stellar initial mass function varies systematically with galaxy mass \citep{Peacock14}. 

The primordial binary fraction is thought to influence the field LMXB population if they form from the evolution of such systems. The small variation observed in \#LMXBs/$\lk$ therefore suggests that the binary fraction must be similar among these galaxies, providing one of the few constraints on binary fractions in unresolved stellar populations. Alternatively, the primordial binary fraction may have little influence on the formation of LMXBs with $\lx>10^{37}~erg/s$. This might suggest an alternative origin of field LMXBs, although we note that a GC origin is unlikely.

\section{Field vs. GC LMXBs} 
\label{sec:xlf_field_gcs}

As noted in Section \ref{sec:xlf}, all of the galaxies observed show hints of differences between the field and GC LMXB populations. To improve our statistics, we now consider the combined populations. To combine these samples together, we must first scale the XLFs for each galaxy. This is because they are associated with different stellar masses/ numbers of GCs and have different X-ray detection limits. We scale each galaxy's XLF using the normalization obtained from fits to the broken power law discussed in Section \ref{sec:xlf_field}. The resulting XLFs are binned into logarithmic bins of width $log(L_{x})=0.2$ and combined together using a weighted mean. 

The combined XLFs of the field LMXBs (grey points) and GC LMXBs (magenta points) are shown in the top panel of Figure \ref{fig:xlf_field_gcs}. The bottom panel shows the ratio of the number of LMXBs in the field to those in GCs. We note that the errors on the XLF are large at high \lx, due to the low number of very high luminosity LMXBs. The errors also increase at low \lx. This is because relatively few of the galaxies in our sample reach these low X-ray detection limits and contribute to this region.

Figure \ref{fig:xlf_field_gcs} suggests that the overall shape of the XLF is different between LMXBs in the field and those in GCs. The GC LMXBs appear to have a flatter distribution, with relatively more LMXBs above $10^{38}erg/s$ and less below. This conclusion is consistent with that of \citet{Zhang11}. A deficit of GC LMXBs with $\lx<5\times10^{37}$~\ergs is also consistent with \citet{Kim09} and \citet{Kim10}. However, we emphasize here that rather than a deficit at any \lx region, the general shape of the XLF appears to be different between these populations - with the GC LMXBs having a flatter XLF. 

To test for the significance of the observed difference between field and GC LMXBs, we chose to combine the individual p-values of each galaxy, rather that run statistics over the combined XLF. We stress that this method has the advantage that it is {\it not} influenced by any errors associated with scaling the XLFs. To combine the p-values we use two different statistics, Stoffer's and Fisher's methods. These two tests suggest the field and GC LMXB populations are significantly different, with similar combined p-values of $(1.9 \text{ and }4.1)\times10^{-6}$ (equivalent to 4.7$\sigma$ for a normal distribution).

\section{Metal-rich vs. metal-poor GC LMXBs}
\label{sec:xlf_gcr_gcb}

We follow the same method described above to scale the red and blue GC LMXB XLFs. In Figure \ref{fig:xlf_field_gcs}, we show the combined metal-rich (red) and metal-poor (blue) GC LMXB XLFs. There are hints in these data that the blue GCs have more LMXBs at high \lx, but this result is not significant. Combining the p-values obtained for each galaxy, we derive p-values for the combined dataset of 0.27 and 0.28 from the Stoffer and Fisher methods, respectively. We therefore conclude that the XLFs of LMXBs in metal-rich and metal-poor GCs are similar. Our conclusions are in agreement with the study of \citet{Kim13}, who showed that the ratio of LMXBs in red and blue GCs (from the ACS Virgo and Fornax cluster surveys) is consistent across this range of \lx. We note that that the red and (especially) blue GC XLFs are relatively uncertain, due to the small number of LMXBs in our sample. Sharper tests of GC metallicity effects will require the addition of data from the GCs of many more galaxies.

\section{Conclusions}
\label{sec:conclusions}

We consider XLF of LMXBs in a sample of seven local, old, early-type galaxies. These galaxies have deep \chandra observations and mosaics of \hst/ACS optical images. Particular attention is given to obtaining clean samples of field and GC LMXBs. We exclude the messy central regions of these galaxies and only include regions covered by \hst images, which are carefully aligned and matched to the X-ray source catalogs to identify GC and background galaxy counterparts. 

We observe a similar number of field LMXBs per stellar mass in these galaxies. This suggests similar formation mechanisms and stellar populations. The XLF of the field LMXBs is significantly different to that of the GC LMXBs, with the latter having a shallower distribution. While there are hints within our data that metal-poor GCs have relatively more LMXBs with $\lx>2\times 10^{38}$~\ergs, we identify no significant difference between the LMXBs in metal-rich and metal-poor clusters. The inclusion of more galaxies will help to improve the statistics and enable sharper tests of metallicity influences in the future. 

\section*{Acknowledgements}

We thank the anonymous referee for providing a prompt and helpful report that benefitted the final version of this paper. 

Support for this work was provided by the NASA through Chandra Award Number AR4-15007X issued by the Chandra X-ray Observatory Center, which is operated by the Smithsonian Astrophysical Observatory for and on behalf of NASA under contract NAS8-03060. Support for this work was provided by NASA through grant number HST-AR-13923 from the Space Telescope Science Institute and from the ADAP grant number NNX15AI71G.

Based on observations made with the NASA/ESA Hubble Space Telescope, and obtained from the Hubble Legacy Archive, which is a collaboration between the Space Telescope Science Institute (STScI/NASA), the Space Telescope European Coordinating Facility (ST-ECF/ESA) and the Canadian Astronomy Data Centre (CADC/NRC/CSA).

The scientific results reported in this article are based in part on data obtained from the Chandra Data Archive and observations made by the Chandra X-ray Observatory and published previously in cited articles.

This research has made use of NASA's Astrophysics Data System.

\bibliographystyle{apj_w_etal}
\bibliography{bibliography_etal}

\begin{thebibliography}{57}
\expandafter\ifx\csname natexlab\endcsname\relax\def\natexlab#1{#1}\fi

\bibitem[{{Angelini} {et~al.}(2001){Angelini}, {Loewenstein}, \&
  {Mushotzky}}]{Angelini01}
{Angelini}, L., {Loewenstein}, M., \& {Mushotzky}, R.~F. 2001, ApJ, 557, L35

\bibitem[{{Ashman} {et~al.}(1994){Ashman}, {Bird}, \& {Zepf}}]{Ashman94}
{Ashman}, K.~M., {Bird}, C.~M., \& {Zepf}, S.~E. 1994, \aj, 108, 2348

\bibitem[{{Bell} \& {de Jong}(2001)}]{Bell01}
{Bell}, E.~F. \& {de Jong}, R.~S. 2001, \apj, 550, 212

\bibitem[{{Bellazzini} {et~al.}(1995){Bellazzini}, {Pasquali}, {Federici},
  {Ferraro}, \& {Pecci}}]{Bellazzini95}
{Bellazzini}, M., {Pasquali}, A., {Federici}, L., {Ferraro}, F.~R., \& {Pecci},
  F.~F. 1995, ApJ, 439, 687

\bibitem[{{Blakeslee} {et~al.}(2009){Blakeslee}, {Jord{\'a}n}, {Mei},
  {C{\^o}t{\'e}}, {Ferrarese}, {Infante}, {Peng}, {Tonry}, \&
  {West}}]{Blakeslee09}
{Blakeslee}, J.~P. {et~al.} 2009, \apj, 694, 556

\bibitem[{{Blakeslee} {et~al.}(2001){Blakeslee}, {Lucey}, {Barris}, {Hudson},
  \& {Tonry}}]{Blakeslee01}
{Blakeslee}, J.~P., {Lucey}, J.~R., {Barris}, B.~J., {Hudson}, M.~J., \&
  {Tonry}, J.~L. 2001, \mnras, 327, 1004

\bibitem[{{Bohlin}(2011)}]{Bohlin11}
{Bohlin}, R.~C. 2011, {Flux Calibration of the ACS CCD Cameras II. Encircled
  Energy Correction}, Tech. rep.

\bibitem[{{Boroson} {et~al.}(2011){Boroson}, {Kim}, \& {Fabbiano}}]{Boroson11}
{Boroson}, B., {Kim}, D.-W., \& {Fabbiano}, G. 2011, \apj, 729, 12

\bibitem[{{Brassington} {et~al.}(2008){Brassington}, {Fabbiano}, {Kim},
  {Zezas}, {Zepf}, {Kundu}, {Angelini}, {Davies}, {Gallagher}, {Kalogera},
  {Fragos}, {King}, {Pellegrini}, \& {Trinchieri}}]{Brassington08}
{Brassington}, N.~J. {et~al.} 2008, \apjs, 179, 142

\bibitem[{{Brassington} {et~al.}(2009){Brassington}, {Fabbiano}, {Kim},
  {Zezas}, {Zepf}, {Kundu}, {Angelini}, {Davies}, {Gallagher}, {Kalogera},
  {Fragos}, {King}, {Pellegrini}, \& {Trinchieri}}]{Brassington09}
---. 2009, \apjs, 181, 605

\bibitem[{{Cappellari} {et~al.}(2012){Cappellari}, {McDermid}, {Alatalo},
  {Blitz}, {Bois}, {Bournaud}, {Bureau}, {Crocker}, {Davies}, {Davis}, {de
  Zeeuw}, {Duc}, {Emsellem}, {Khochfar}, {Krajnovi{\'c}}, {Kuntschner},
  {Lablanche}, {Morganti}, {Naab}, {Oosterloo}, {Sarzi}, {Scott}, {Serra},
  {Weijmans}, \& {Young}}]{Cappellari12}
{Cappellari}, M. {et~al.} 2012, \nat, 484, 485

\bibitem[{{Clark}(1975)}]{Clark75}
{Clark}, G.~W. 1975, ApJ, 199, L143

\bibitem[{{de Vaucouleurs} {et~al.}(1991){de Vaucouleurs}, {de Vaucouleurs},
  {Corwin}, {Buta}, {Paturel}, \& {Fouqu{\'e}}}]{deVaucouleurs91}
{de Vaucouleurs}, G., {de Vaucouleurs}, A., {Corwin}, Jr., H.~G., {Buta},
  R.~J., {Paturel}, G., \& {Fouqu{\'e}}, P. 1991, {Third Reference Catalogue of
  Bright Galaxies.} (Springer, New York, USA)

\bibitem[{{Fall} \& {Romanowsky}(2013)}]{Fall13}
{Fall}, S.~M. \& {Romanowsky}, A.~J. 2013, \apjl, 769, L26

\bibitem[{{Fragos} {et~al.}(2008){Fragos}, {Kalogera}, {Belczynski},
  {Fabbiano}, {Kim}, {Brassington}, {Angelini}, {Davies}, {Gallagher}, {King},
  {Pellegrini}, {Trinchieri}, {Zepf}, {Kundu}, \& {Zezas}}]{Fragos08}
{Fragos}, T. {et~al.} 2008, \apj, 683, 346

\bibitem[{{Grindlay}(1984)}]{Grindlay84}
{Grindlay}, J.~E. 1984, Advances in Space Research, 3, 19

\bibitem[{{Grindlay}(1993)}]{Grindlay93}
{Grindlay}, J.~E. 1993, in Astronomical Society of the Pacific Conference
  Series, Vol.~48, The Globular Cluster-Galaxy Connection, ed. G.~H. {Smith} \&
  J.~P. {Brodie}, 156

\bibitem[{{Humphrey} \& {Buote}(2008)}]{Humphrey08}
{Humphrey}, P.~J. \& {Buote}, D.~A. 2008, \apj, 689, 983

\bibitem[{{Ivanova} {et~al.}(2013){Ivanova}, {Justham}, {Chen}, {De Marco},
  {Fryer}, {Gaburov}, {Ge}, {Glebbeek}, {Han}, {Li}, {Lu}, {Marsh},
  {Podsiadlowski}, {Potter}, {Soker}, {Taam}, {Tauris}, {van den Heuvel}, \&
  {Webbink}}]{Ivanova13}
{Ivanova}, N. {et~al.} 2013, \aapr, 21, 59

\bibitem[{{Jardel} {et~al.}(2011){Jardel}, {Gebhardt}, {Shen}, {Fisher},
  {Kormendy}, {Kinzler}, {Lauer}, {Richstone}, \& {G{\"u}ltekin}}]{Jardel11}
{Jardel}, J.~R. {et~al.} 2011, \apj, 739, 21

\bibitem[{{Jarrett} {et~al.}(2003){Jarrett}, {Chester}, {Cutri}, {Schneider},
  \& {Huchra}}]{Jarrett03}
{Jarrett}, T.~H., {Chester}, T., {Cutri}, R., {Schneider}, S.~E., \& {Huchra},
  J.~P. 2003, \aj, 125, 525

\bibitem[{{Jensen} {et~al.}(2003){Jensen}, {Tonry}, {Barris}, {Thompson},
  {Liu}, {Rieke}, {Ajhar}, \& {Blakeslee}}]{Jensen03}
{Jensen}, J.~B. {et~al.} 2003, \apj, 583, 712

\bibitem[{{Jord{\'a}n} {et~al.}(2004)}]{Jordan04}
{Jord{\'a}n}, A. {et~al.} 2004, ApJ, 613, 279

\bibitem[{{Jord{\'a}n} {et~al.}(2007)}]{Jordan07}
---. 2007, ApJ, 671, L117

\bibitem[{Joseph(2013)}]{Joseph13}
Joseph, T. 2013, PhD thesis, University of Southampton

\bibitem[{{Kim} \& {Fabbiano}(2004)}]{Kim04}
{Kim}, D. \& {Fabbiano}, G. 2004, ApJ, 611, 846

\bibitem[{{Kim} \& {Fabbiano}(2010)}]{Kim10}
{Kim}, D.-W. \& {Fabbiano}, G. 2010, \apj, 721, 1523

\bibitem[{{Kim} {et~al.}(2009){Kim}, {Fabbiano}, {Brassington}, {Fragos},
  {Kalogera}, {Zezas}, {Jord{\'a}n}, {Sivakoff}, {Kundu}, {Zepf}, {Angelini},
  {Davies}, {Gallagher}, {Juett}, {King}, {Pellegrini}, {Sarazin}, \&
  {Trinchieri}}]{Kim09}
{Kim}, D.-W. {et~al.} 2009, \apj, 703, 829

\bibitem[{{Kim} {et~al.}(2013){Kim}, {Fabbiano}, {Ivanova}, {Fragos},
  {Jord{\'a}n}, {Sivakoff}, \& {Voss}}]{Kim13}
{Kim}, D.-W., {Fabbiano}, G., {Ivanova}, N., {Fragos}, T., {Jord{\'a}n}, A.,
  {Sivakoff}, G.~R., \& {Voss}, R. 2013, \apj, 764, 98

\bibitem[{{Kim} {et~al.}(2007){Kim}, {Lee}, {Geisler}, {Sarajedini}, {Park},
  {Hwang}, {Harris}, {Seguel}, \& {von Hippel}}]{Kim07}
{Kim}, S.~C. {et~al.} 2007, AJ, 134, 706

\bibitem[{{Kundu} {et~al.}(2002){Kundu}, {Maccarone}, \& {Zepf}}]{Kundu02}
{Kundu}, A., {Maccarone}, T.~J., \& {Zepf}, S.~E. 2002, ApJ, 574, L5

\bibitem[{{Kundu} {et~al.}(2007){Kundu}, {Maccarone}, \& {Zepf}}]{Kundu07}
---. 2007, ApJ, 662, 525

\bibitem[{{Kundu} \& {Whitmore}(2001)}]{Kundu01}
{Kundu}, A. \& {Whitmore}, B.~C. 2001, \aj, 122, 1251

\bibitem[{{Lehmer} {et~al.}(2014){Lehmer}, {Berkeley}, {Zezas}, {Alexander},
  {Basu-Zych}, {Bauer}, {Brandt}, {Fragos}, {Hornschemeier}, {Kalogera},
  {Ptak}, {Sivakoff}, {Tzanavaris}, \& {Yukita}}]{Lehmer14}
{Lehmer}, B.~D. {et~al.} 2014, \apj, 789, 52

\bibitem[{{Li} {et~al.}(2010){Li}, {Spitler}, {Jones}, {Forman}, {Kraft}, {Di
  Stefano}, {Tang}, {Wang}, {Gilfanov}, \& {Revnivtsev}}]{Li10}
{Li}, Z. {et~al.} 2010, \apj, 721, 1368

\bibitem[{{Liu}(2011)}]{Liu11}
{Liu}, J. 2011, \apjs, 192, 10

\bibitem[{{Liu} {et~al.}(2001){Liu}, {van Paradijs}, \& {van den
  Heuvel}}]{Liu01}
{Liu}, Q.~Z., {van Paradijs}, J., \& {van den Heuvel}, E.~P.~J. 2001, A\&A,
  368, 1021

\bibitem[{{Luo} {et~al.}(2013){Luo}, {Fabbiano}, {Strader}, {Kim}, {Brodie},
  {Fragos}, {Gallagher}, {King}, \& {Zezas}}]{Luo13}
{Luo}, B. {et~al.} 2013, \apjs, 204, 14

\bibitem[{{Mineo} {et~al.}(2014){Mineo}, {Fabbiano}, {D'Abrusco}, {Fragos},
  {Kim}, {Strader}, {Brodie}, {Gallagher}, {Zezas}, \& {Luo}}]{Mineo14}
{Mineo}, S. {et~al.} 2014, \apj, 780, 132

\bibitem[{{Muratov} \& {Gnedin}(2010)}]{Muratov10}
{Muratov}, A.~L. \& {Gnedin}, O.~Y. 2010, \apj, 718, 1266

\bibitem[{{Paolillo} {et~al.}(2011){Paolillo}, {Puzia}, {Goudfrooij}, {Zepf},
  {Maccarone}, {Kundu}, {Fabbiano}, \& {Angelini}}]{Paolillo11}
{Paolillo}, M. {et~al.} 2011, \apj, 736, 90

\bibitem[{{Peacock} {et~al.}(2010){Peacock}, {Maccarone}, {Kundu}, \&
  {Zepf}}]{Peacock10b}
{Peacock}, M.~B., {Maccarone}, T.~J., {Kundu}, A., \& {Zepf}, S.~E. 2010,
  \mnras, 407, 2611

\bibitem[{{Peacock} {et~al.}(2009){Peacock}, {Maccarone}, {Waters}, {Kundu},
  {Zepf}, {Knigge}, \& {Zurek}}]{Peacock09}
{Peacock}, M.~B., {Maccarone}, T.~J., {Waters}, C.~Z., {Kundu}, A., {Zepf},
  S.~E., {Knigge}, C., \& {Zurek}, D.~R. 2009, MNRAS, 392, L55

\bibitem[{{Peacock} {et~al.}(2014){Peacock}, {Zepf}, {Maccarone}, {Kundu},
  {Gonzalez}, {Lehmer}, \& {Maraston}}]{Peacock14}
{Peacock}, M.~B., {Zepf}, S.~E., {Maccarone}, T.~J., {Kundu}, A., {Gonzalez},
  A.~H., {Lehmer}, B.~D., \& {Maraston}, C. 2014, \apj, 784, 162

\bibitem[{{Pooley} {et~al.}(2003){Pooley}, {Lewin}, {Anderson}, {Baumgardt},
  {Filippenko}, {Gaensler}, {Homer}, {Hut}, {Kaspi}, {Makino}, {Margon},
  {McMillan}, {Portegies Zwart}, {van der Klis}, \& {Verbunt}}]{Pooley03}
{Pooley}, D. {et~al.} 2003, ApJ, 591, L131

\bibitem[{{Rhode} \& {Zepf}(2001)}]{Rhode01}
{Rhode}, K.~L. \& {Zepf}, S.~E. 2001, \aj, 121, 210

\bibitem[{{Saglia} {et~al.}(2000){Saglia}, {Kronawitter}, {Gerhard}, \&
  {Bender}}]{Saglia00}
{Saglia}, R.~P., {Kronawitter}, A., {Gerhard}, O., \& {Bender}, R. 2000, \aj,
  119, 153

\bibitem[{{S{\'a}nchez-Bl{\'a}zquez} {et~al.}(2006){S{\'a}nchez-Bl{\'a}zquez},
  {Gorgas}, {Cardiel}, \& {Gonz{\'a}lez}}]{Sanchez-Blazquez06b}
{S{\'a}nchez-Bl{\'a}zquez}, P., {Gorgas}, J., {Cardiel}, N., \& {Gonz{\'a}lez},
  J.~J. 2006, \aap, 457, 809

\bibitem[{{Sivakoff} {et~al.}(2008){Sivakoff}, {Jord{\'a}n}, {Juett},
  {Sarazin}, \& {Irwin}}]{Sivakoff08}
{Sivakoff}, G.~R., {Jord{\'a}n}, A., {Juett}, A.~M., {Sarazin}, C.~L., \&
  {Irwin}, J.~A. 2008, ArXiv e-prints

\bibitem[{{Sivakoff} {et~al.}(2007){Sivakoff}, {Jord{\'a}n}, {Sarazin},
  {Blakeslee}, {C{\^o}t{\'e}}, {Ferrarese}, {Juett}, {Mei}, \&
  {Peng}}]{Sivakoff07}
{Sivakoff}, G.~R. {et~al.} 2007, \apj, 660, 1246

\bibitem[{{Taylor}(2006)}]{Taylor06}
{Taylor}, M.~B. 2006, in Astronomical Society of the Pacific Conference Series,
  Vol. 351, Astronomical Data Analysis Software and Systems XV, ed.
  C.~{Gabriel}, C.~{Arviset}, D.~{Ponz}, \& S.~{Enrique}, 666

\bibitem[{{Thomas} {et~al.}(2010){Thomas}, {Maraston}, {Schawinski}, {Sarzi},
  \& {Silk}}]{Thomas10}
{Thomas}, D., {Maraston}, C., {Schawinski}, K., {Sarzi}, M., \& {Silk}, J.
  2010, \mnras, 404, 1775

\bibitem[{{Verbunt} \& {Hut}(1987)}]{Verbunt87}
{Verbunt}, F. \& {Hut}, P. 1987, in IAU Symposium, Vol. 125, The Origin and
  Evolution of Neutron Stars, ed. {D.~J.~Helfand \& J.-H.~Huang}, 187--+

\bibitem[{{Villegas} {et~al.}(2010){Villegas}, {Jord{\'a}n}, {Peng},
  {Blakeslee}, {C{\^o}t{\'e}}, {Ferrarese}, {Kissler-Patig}, {Mei}, {Infante},
  {Tonry}, \& {West}}]{Villegas10}
{Villegas}, D. {et~al.} 2010, \apj, 717, 603

\bibitem[{{Voss} {et~al.}(2009){Voss}, {Gilfanov}, {Sivakoff}, {Kraft},
  {Jord{\'a}n}, {Raychaudhury}, {Birkinshaw}, {Brassington}, {Croston},
  {Evans}, {Forman}, {Hardcastle}, {Harris}, {Jones}, {Juett}, {Murray},
  {Sarazin}, {Woodley}, \& {Worrall}}]{Voss09}
{Voss}, R. {et~al.} 2009, \apj, 701, 471

\bibitem[{{Zhang} {et~al.}(2012){Zhang}, {Gilfanov}, \& {Bogd{\'a}n}}]{Zhang12}
{Zhang}, Z., {Gilfanov}, M., \& {Bogd{\'a}n}, {\'A}. 2012, \aap, 546, A36

\bibitem[{{Zhang} {et~al.}(2011){Zhang}, {Gilfanov}, {Voss}, {Sivakoff},
  {Kraft}, {Brassington}, {Kundu}, {Jord{\'a}n}, \& {Sarazin}}]{Zhang11}
{Zhang}, Z. {et~al.} 2011, \aap, 533, A33

\end{thebibliography}

\label{lastpage}

\end{document}